\documentclass[useAMS,usegraphicx]{mn2e}

\def\lesssim{\,\lower2truept\hbox{${<\atop\hbox{\raise4truept\hbox{$\sim$}}}$}\,}
\def\gtrsim{\,\lower2truept\hbox{${>\atop\hbox{\raise4truept\hbox{$\sim$}}}$}\,}

\def\rev{}

\title[Puffing up ETGs: numerical experiments]{Puffing up early-type galaxies by baryonic mass
loss: \\ numerical experiments}

\author[Ragone-Figueroa \& Granato]{
\parbox[t]{\textwidth}{
Cinthia Ragone-Figueroa$^1$\thanks{Email: cin@mail.oac.uncor.edu},
Gian~Luigi Granato$^{2}$\thanks{Email: granato@oats.inaf.it}}
\vspace*{6pt} \\
  $^1$ Instituto de Astronom\'ia Te\'orica y Experimental, IATE, CONICET-Observatorio Astron\'omico, \\
  Universidad Nacional de C\'ordoba, Laprida 854, X5000BGR, C\'ordoba, Argentina\\
  $^2$ Istituto Nazionale di Astrofisica INAF, Osservatorio Astronomico di Trieste, Via Tiepolo 11, I-34131
  Trieste, Italy \\
}

\begin{document}
\date{Accepted ... Received ...}

\maketitle

\begin{abstract}

{\rev Observations performed in the last few years indicate that most
massive early-type galaxies (ETGs) observed at redshift $z\ga 1$ exhibit
sizes smaller by a factor of a few than local ETGs of analogous stellar
mass.} We present numerical simulations of the effect of baryonic mass
loss on the structure of a spheroidal stellar system, embedded in a dark
matter halo. This process, {\rev invoked as a possible explanation of the
observed size increase of ETGs since $z \sim 2$,} could be caused either
by QSO/starburst driven galactic winds, promptly ejecting from Early Type
Galaxies (ETGs) the residual gas {\rev and halting star formation}
(galactic winds), or by stellar mass returned to the ISM in the final
stages of stellar evolution. {\rev Indeed}, we find that a conceivable
loss of $\sim 50 \%$ of the baryonic mass can produce a significant size
increase. However, the puffing up {\rev due to  galactic winds} occurs
when the stellar populations are much younger than the estimated ages
$\gtrsim 0.5$ Gyr of compact high-z ETGs. {\rev Therefore, while it may
have had a role in deciding the final structure of ETGs, it cannot explain
the evolution observed so far of their size-mass relation; its signature
should be searched for in much younger systems.} Conversely, the mass loss
due to stellar evolution {\rev could cause a relatively modest expansion
of passively evolving stellar systems later on,} contributing to, without
dominating, the observed evolution of their mass-size relationship.
\end{abstract}


\section{Introduction}

During the last years it has been established that most massive early-type
galaxies (ETGs) observed at redshift $z\ga 1$ exhibit sizes smaller by a
factor of a few than local ETGs of analogous stellar mass (e.g.\ {\rev
Daddi et al.\ 2004; Trujillo et al.\ 2006, 2007, 2011;} Longhetti et al.\
2007; Toft et al.\ 2007; Zirm et al.\ 2007; van der Wel et al.\ 2008; van
Dokkum et al.\ 2008; Cimatti et al.\ 2008; Buitrago et al.\ 2008; Damjanov
et al.\ 2009; Ryan et al.\ 2010).

The possibility that the size evolution is, at least in part, apparent and
due to some subtle systematic effect has not been completely ruled out.
Discussed caveats include a centrally concentrated source (such as an AGN
or central starburst), age gradients, un-detected low surface brightness
external regions at high z, or a top-heavy IMF affecting mass estimates
(see Daddi et al.\ 2005; Van Dokkum et al.\ 2008; La Barbera et al.\ 2009;
Mancini et al. 2010; Hopkins et al.\ 2009, 2010). {\rev Moreover, very
recent observational results claim some co-existence of compact and normal
size ETGs, both at high (Mancini et al.\ 2010; Onodera et al.\ 2010;
Saracco et al.\ 2010) and low redshift (Valentinuzzi et al.\ 2010).}

{\rev However, nowadays most authors agree that the observational results
are dominated by a real size evolution.} The proposed interpretations are
related either to the effects of mergers or to the loss of a substantial
fraction of mass from the galaxy.

Khochfar \& Silk (2006) presented a semi-analytic model where the size
evolution is substantial only for galaxies more massive than $10^{11}$
M$_\odot$, and results from massive galaxies at high redshifts forming in
gas-rich dissipative mergers, whilst galaxies of the same mass at low
redshifts form from gas-poor mergers. However, current observations
indicate sizeable evolution also in much smaller systems (e.g.\ Ryan et al
2010). Similar ideas have been explored by Hopkins et al.\ (2009), in a
more phenomenological model incorporating results of a large suite of
numerical simulations of mergers.

Actually, the size increasing effects of {\it major} (wherein the two
merging galaxies have comparable masses) {\it dry} (i.e.\ without a
significant collisional gas component) mergers has been often discussed.
By converse, in {\it wet} mergers, the presence of a dissipative gas
component limits the gain in size. However, even the former process faces
a few problems. The most basic is that it would move galaxies too slowly
toward the local size-mass relationship ({\rev Cimatti et al 2008;}
Bezanson et al.\ 2009; Damjanov et al.\ 2009; {\rev Saracco et al 2009}).
This is because, according to simple virial theorem argument, confirmed by
a number of numerical simulations, in major dry mergers the size increases
almost linearly with the mass, and possibly somewhat less (e.g.\ Ciotti \&
van Albada 2001; Nipoti  et  al.\ 2003; Boylan-Kolchin et al. 2006; Naab
et al.\ 2009). This dependence is too close to the observed local
mass-size relationship $r \propto M^{0.56}$ (Shen et al.\ 2003) to explain
the evolution in a reasonable number of merging events.

The most promising {\it merging} mechanism to explain the size increase
seems to be a series of late minor dry merging events. These would add
stars in the outer parts of passive high-z galaxies, in such a way to
produce a size increase that can scale as steep as $M^2$ (e.g. Naab et al.
2009; Oser et al.\ 2010). Actually, Hopkins et al.\ (2010), considering
the various proposed channels for observed size evolution, concluded that
minor dry mergers are the best candidate to dominate, though other
channels should have a non negligible role.

In this paper, we will test the possible contribution of the {\it
puffing-up} process. This envisages that the expansion in size is driven
by the expulsion of a substantial fraction of the gas out of the galaxy
either by AGN and/or supernova driven galactic winds (Fan et al.\ 2008,
2010), or by the expulsion of gas associated to stellar evolution
(Damjanov et al. 2009). In the former case the expulsion timescale would
be short, likely not much longer than the dynamical timescale, at least
when driven by the AGN, whilst in the latter an important mass loss could
last even $\sim 0.5-1$ Gyr, depending on the IMF and on stellar evolution
details.

While the works by Fan et al.\ (2008, 2010) used as a reference the
specific semi-analytic model by Granato et al.\ (2004) for the
co-evolution of ETGs and SMBH (henceforth G04), this kind of puffing-up
due to baryonic mass loss from the galaxy has a broader applicability.
{\rev Indeed, virtually all modern models of galaxy formation give a
prominent role to AGN and/or SNae driven galactic wind, ejecting from the
galactic region a substantial fraction of gas (for a recent review, see
Benson 2010).} Thus, it is {\rev very likely} that this puffing up played
a role in deciding the final sizes of ETGs, at some point over their
history. However, it is still an open question whether this role has been
major, and, in particular, whether it can explain the available
observations. The aim of the present work is to provide a step to clarify
it.

{\rev The idea that a puffing-up mechanism should at some point have had
occurred} rests on many hints, nowadays coming from a complex interplay
between observations and theory, suggesting that ETGs are the result of an
intense phase of high-z star formation activity, possibly traced by the
sub-mm selected galaxy population. It is unlikely that these extreme
"starburst", induced by fast collapse and/or by rapid merging of gas rich
systems, have been suddenly terminated by simple gas consumption. It is
instead usually envisaged that some strong feedback was capable to eject
from the galaxy, on a relatively short timescale, a substantial fraction
of its baryonic mass, in the form of gas not yet converted into stars (a
process usually referred to as ''galactic wind or superwind''; see e.g.\
Benson et al.\ 2003;  Pipino, Silk \& Matteucci 2009; for observational
indications of these high-z galactic outflows see L\'{\i}pari et al.\ 2009
and Prochaska \& Hennawy 2009, and references therein). In the past years
it has been pointed out that the most likely candidate to power such a
process, at least in massive systems, is QSO activity (e.g.\ Silk \& Rees
1998; Fabian 1999; Granato et al.\ 2001, 2004; Benson et al 2003; Cattaneo
et al. 2006; Monaco et al.\ 2007; Sijacki et al.\ 2007; Somerville et al.\
2008; Johansson et al.\ 2009; Ciotti, Ostriker \& Proga 2009).

Additionally, after the termination of this huge star-forming phase, it is
likely that the galaxy lost another significant fraction of its baryonic
mass, due to stellar evolution (supernova explosions and stellar winds).

Independently of the still uncertain details of the formation mechanism of
ETGs, or more generally of the spheroidal component of galaxies, it seems
timely to investigate with aimed numerical simulations the effects of the
expulsion of a significant fraction of baryonic mass from a spheroidal
system embedded in a dark matter (DM) halo.

The plan of the paper is {\rev the following. In Section \ref{sec:cluster}
we recall previous results on a problem sharing similarities with that
considered here: the dynamical evolution of young star clusters after
dispersion of the parent gas cloud. These results inspired the proposal
that the size evolution of ETGs could be due to the puffing up mechanism,
and were adopted as a rough approximation for quantitative
considerations.} Section \ref{sec:method} we describe the simulation
technique and the initial conditions, the results are presented in Section
\ref{sec:results}, and discussed, in the context of observed size
evolution of ETGs, in the final Section \ref{sec:discussion}. We use the
concordance cosmology (Komatsu et al.\ 2009), i.e., a flat universe with
matter density parameter $\Omega_{M} = 0.3$ and Hubble constant $H_0 = 70$
km s$^{-1}$ Mpc$^{-1}$.

\section{The star cluster approximation}
\label{sec:cluster}

{\rev A process somewhat similar to the puffing up of ETGs by mass loss}
has been addressed several times, both analytically and numerically, in
the context of the dynamical evolution of star clusters. Again, the mass
loss is due to (i) the indirect and combined effects of stellar winds and
supernova explosions, soon driving out the significant fraction of gas not
used in star formation, or (ii) directly to stellar evolution. As for star
clusters, the latter contribution includes the gas ejected from the stars
by winds and explosions, but also the compact renmants that may get at
birth a kick velocity sufficient to be ejected from the system. The most
obvious differences, with respect to the puffing-up of ETGs, are the
absence of an embedding dark matter halo and, to a lesser extent, the
importance of two-body collisions.

Biermann \& Shapiro (1979) and Hills (1980), under simplifying assumptions
to allow an analytical treatment, found that the size evolution depends on
the ejection timescale. If this is short compared to the dynamical time
(hereafter {\it fast ejection}), conservation of specific kinetic energy
yields for the expansion factor
\begin{equation}
{R_{\rm f}\over R_{\rm i}}={\epsilon\over 2\epsilon-1}
\end{equation}
in terms of the ratio between the final and initial mass $\epsilon\equiv
M_{\rm f}/M_{\rm i}$; note that for $\epsilon\leq 0.5$ the system becomes
unbound and dissociates. On the other hand, if the ejection timescales is
much longer than the dynamical one, conservation of adiabatic invariants
yields the size evolution
\begin{equation}
{R_{\rm f}\over R_{\rm i}}={1\over \epsilon}~;
\end{equation}
Thus, fast expulsion is more effective in increasing the size, while, if
the expulsion is slow enough (adiabatic), the system remains bound
independently of $\epsilon$.

These relationships have been checked and substantially confirmed by
several numerical simulations of star clusters dynamics, starting from the
pioneristic work by Tutukov (1978). However, due to a few effects not
accounted for in analytical works, a portion of the system remains bound
even if $\epsilon$ is somewhat smaller than 0.5 and the mass loss is fast
(see Baumgardt \& Kroupa 2007 and references therein). Numerical
experiments show also that, after fast mass loss ends, the system rapidly
reaches a maximum transient expansion within $10-15$ dynamical times,
while a new equilibrium is attained over $30-40$ dynamical times (e.g.\
Geyer \& Bukert, 2001; Goodwin \& Bastian, 2006; Baumgardt \& Kroupa,
2007).

Fan et al.\ (2008, 2010) tentatively tested against available data a QSO
driven puffing up scenario for ETGs, resting on the G04 model of joint
evolution of SMBH and spheroids, and adopting the above relationships
coming from star cluster dynamics. Their findings are to some extent
encouraging, but, {\rev as remarked by Mancini et al.\ (2010),} a closer
analysis reveals that the expansion time-scale seems far too short to
explain the relatively old ($\gtrsim 1$ Gyr) stellar ages claimed for
high-z compact galaxies. A similar, albeit less dramatic, problem  has
been pointed out by Damjanov et al.\ (2009) when trying to explain the
observed size evolution by means of mass loss due to stellar evolution. In
this case the timescale of mass loss are dictated by stellar evolution,
and the expected expansion is adiabatic.

However, assuming the above recipes for ETGs is only a zeroth order
approximation, since in ETGs the DM halo is expected to affect the
efficiency and timescale of the size evolution, and to prevent galaxy
disruption even when a major fraction of baryonic mass is lost. {\rev In
the following,} our purpose is to investigate these effects via simple but
aimed numerical simulations.

\begin{figure}
  \centerline{\includegraphics[width=9cm]{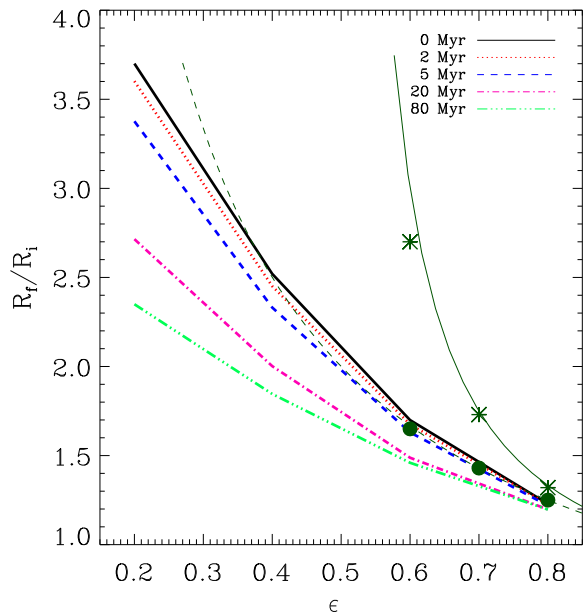}}
  \caption{Half mass radius increase $R_{\rm f}/R_{\rm i}$ due to baryonic mass loss as
  a function of the parameter $\epsilon\equiv
M_{B,fin}/M_{B,ini}$, for different ejection times $\Delta t$ {\rev (thick lines are with DM included)}.
The thin lines illustrate the size
increase according to the analytic approximation, {\rev meant for a system not embedded in a DM halo}, of Eq.~(1) (solid, fast ejection) and (2) (dashed
slow ejection).
The points show examples of runs without DM halo included (asterisks for fast ejection $\Delta t=0$ and
circles for slow ejection $\Delta t=80$ Myr)
} \label{fig:sizinc}
\end{figure}

\begin{figure*}
  \centerline{\includegraphics[width=18cm]{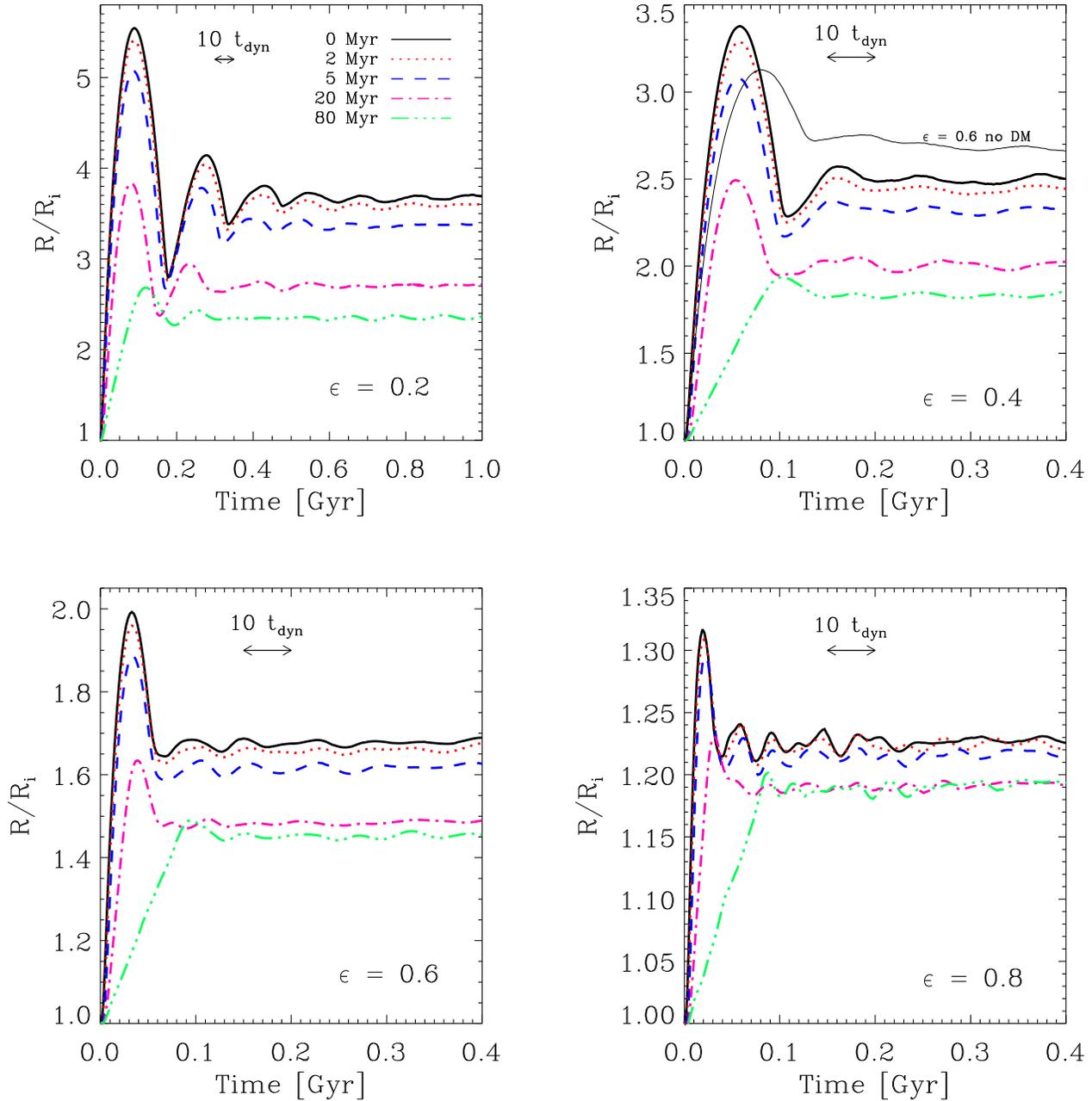}}
  \caption{Ratio $R/R_{\rm i}$ of the current to the initial half-mass radius as
a function of time, for different values of the diet parameter $\epsilon$
and of the ejection times $\Delta t$, as indicated in the panels. The thin
solid line in the top right panel shows the evolution of a model not
including the DM halo, for $\epsilon=0.6$. {\rev Note that the latter is
shown in the panel corresponding to $\epsilon=0.4$ (for the cases
including DM) since the vertical scale is more adequate}. The double
arrows show the duration of a 10 $t_{\rm dyn}$ time interval. For the
adopted initial conditions, $t_{\rm dyn} \simeq 5$ Myr. See text for
explanations.} \label{fig:rad_t}
\end{figure*}

\section{Numerical method and setup}
\label{sec:method}

The purpose of the simulations is to investigate the evolution of
collision-less particles (stars and DM) under a change of gravitational
potential due to a loss of baryonic mass of the system. The escaping mass
can be either the gas which has not been converted into stars during the
star forming phase of the spheroid, or the mass lost from stars in form of
stellar winds and SNae explosions. In any case, we assume as given, and
due to ``external'' causes (such as SNae and AGN feedbacks, or stellar
evolution), the temporal dependence of this mass loss
(Eq.~\ref{eq:massloss}), which we put by hand, and we simulate the ensuing
evolution of collision-less mass distributions. Therefore we don't have to
treat the gas dynamics. This is the same approach followed in most
simulations of puffing-up of star clusters (e.g.\ Boily \& Kroupa, 2003;
Goodwin \& Bastian, 2006; Baumgardt \& Kroupa, 2007).

We used the $N-$body code \textsl{GADGET-2} (Springel et al.\ 2005) to
perform simulations with $10^6$ and $5\times 10^{6}$ particles. {\rev None
of the presented results show any noticeable difference in the two cases,
which assures us that the mass resolution is sufficient for the purposes
of the present study.} Half of the particles are used to sample the
baryonic and dark matter components respectively.

The density distribution of DM particles is assumed to follow the standard
NFW (Navarro, Frenk \& White 1997) shape
\begin{equation}
\rho_{\rm DM}(r)={M_{\rm vir, DM}\over 4\pi\, R_{\rm
vir}^3}\,{c^2\,g(c)\over \hat r\, (1+c\, \hat r)^2}~,
\label{eq:nfw}
\end{equation}
where $M_{\rm vir, DM}$ is the halo virial mass in DM (the DM mass inside
$R_{\rm vir}$), $\hat r= r/R_{\rm vir}$, $c$ is the concentration
parameter and $g(c)\equiv [\log(1+c)-c/(1+c)]^{-1}$.


The virial radius, {\rev $R_{\rm vir},$ is by definition} the radius within which
the mean density is $\Delta_{\rm vir}(z)$ (a quantity
classically coming from the spherical top-hat collapse model)
times the mean matter density of the universe $\rho_{u}(z)$:
\begin{equation}
R_{\rm vir}=\left[\frac{3}{4 \pi} \frac{M_{\rm vir}}{\Delta_{\rm vir}(z)\rho_{u}(z)}
\right]^{1/3}, \label{Rvir}
\end{equation}

{\rev The overdensity $\Delta_{\rm vir}(z)$, for a flat cosmology, can be
approximated by}

\begin{equation}
\Delta_{\rm vir}(z) \simeq \frac{(18 \pi^2 +82x-39x^2)}{\Omega(z)},
\label{overdensity}
\end{equation}

{\rev where $x= \Omega(z)-1$ and $\Omega(z)$ is the ratio of the mean
matter density to the critical density at redshift $z$ (Bryan \& Norman
1998).}

The corresponding mass distribution is written
\begin{equation}
M_{\rm DM}(<r)=M_{\rm vir,  DM}\, g(c)\,
\left[\log{(1+c \, \hat r)}-{c \, \hat r\over
(1+c\, \hat r)}\right]~.
\end{equation}
To cope with the divergence of the NFW mass distribution, we introduce an
exponential cut at $3 R_{\rm vir}$, {\rev which very safely contains the
region where the baryonic component is important. We performed also a few
test runs with the cut at $2 R_{\rm vir}$, verifying that none of the
discussed quantities is affected by this choice.}

{\rev For} the baryonic particles (stars and gas), we assume that they follow
an Hernquist (1990) profile, which provides a reasonable description of
stellar density in local spheroids:
\begin{equation}
\rho_{\rm B}(r)={M_{\rm B}\over 2 \pi } {a \over r} {1 \over (r+a)^3}~;
\label{eq:her}
\end{equation}
The corresponding mass distribution is
\begin{equation}
M_{\rm B}(<r)=M_{\rm B}\, \left({ r \over r+a}\right)^2~;
\end{equation}
so that the half-mass radius is related to the scale radius $a$ by
$R_{1/2}=(1+\sqrt{2})\, a$ and, assuming a mass {\rev to} light ratio
independent of $r$, the effective radius is $R_{e}\simeq 1.81 a$.

In the following, unless otherwise specified, by {\it dynamical time}
$t_{\rm dyn}$ we mean the initial (i.e.\ before any mass loss and
expansion) dynamical time, computed at $R_{1/2}$.
\begin{equation}
t_{\rm dyn}=\left[R_{1/2}^3\over 2 G \left(M_{\rm B}/2+M_{\rm DM}(<R_{1/2})\right) \right]^{1/2}
\label{eq:tdyn}
\end{equation}
For our standard initial conditions (see below) the contribution of DM to
the mass inside $R_{1/2}$ amounts to $\simeq 20$\%. Thus $t_{\rm dyn}$ can
be estimated (within 10\%) neglecting it:
\begin{equation}
t_{\rm dyn}\simeq 2.3 \left(R_e\over 1 {\rm kpc}\right)^{1.5} \left(M_{\rm B}\over 10^{11}
M_\odot\right)^{-0.5}
\,\, {\rm Myr}
\label{eq:tdynapp}
\end{equation}

Given the density runs, we obtain the 1D velocity dispersion by
integrating the Jeans equation under the assumption of isotropic
conditions:
\begin{equation}
\sigma_{X}^2(r)=-{1\over \rho_{X}(r)}\,\int_r^{\infty}{\rm
d}{r'}\, {G\, M_{\rm TOT}(<r')\over
r'^2}\, \rho_{X}(r')~,
\end{equation}
where $X$ stands for B or DM, and $M_{\rm TOT}(<r)=M_{\rm DM}(<r)+M_{\rm
B}(<r)$. By evolving the particle system for several dynamical times, we
get confident that it is actually in (quasi-)static statistical
equilibrium.

Starting from this initial setup, we introduce a mass loss, intended to
emulate the various possible effects described above, by removing
exponentially over an ejection time $\Delta t$ a fraction $1-\epsilon$ of
the baryonic mass :
\begin{equation}
M_B(t)=M_{\rm B(t=0)}\,\exp\left({\ln \epsilon \over \Delta t} \, t\right)~,
\label{eq:massloss}
\end{equation}
For instance, this simple functional form provides an acceptable
description of the gas removal due to QSO feedback in the G04
semi-analytic model, with an ejection timescale of the order of 20-30 Myr
for a wide range of the model parameters.

The mass loss is practically attained by decreasing correspondingly in
time the mass of the baryonic particles sampling the density field. After
the end of the mass loss period, we let the system to evolve till it
reaches, if any, a new equilibrium configuration.

The reference value for the initial (i.e.\ before any mass loss) ratio of
virial mass (total mass within the virial radius) to baryonic mass is
$M_{\rm vir}/M_{\rm B(t=0)}=25$. In a recent analysis consistent with
previous works, Monster et al.\ 2010  find that this ratio should be, in
the local universe, about 50 for DM haloes of $\sim 5 \times 10^{12}$
M$_{\odot}$.  However, to get a significant puffing-up, the system should
lose something of the order of 50\% of its baryonic mass. Thus, we set as
initial condition a ratio twice smaller than that.

{\rev We set $M_{\rm vir}=10^{13}{\rm M}_\odot$ in all simulations.
Nevertheless, our results apply to different values of $M_{\rm vir}$,
provided that the ratios of scale radii and masses in the two components
(DM and baryons) are not changed, and that the time is measured in units
of dynamical time $t_{\rm dyn} \propto\rho^{-1/2}$.}

We adopt  a concentration parameter $c=4$, a typical value at galactic
halo formation (see Zhao et al. 2003; Klypin, Trujillo-Gomez, \& Primack
2010), and $R_{\rm vir}\simeq 170$ kpc, which is the value given by
Eq.~\ref{Rvir} and Eq.~\ref{overdensity}, for a $M_{\rm vir}=10^{13} {\rm
M}_\odot$ halo virialized at $z=3$.

We set $a=1.5$ kpc ($R_e\simeq 2.7$ kpc). This seems a value adequate to
study the evolution of the system in the plane effective radius $R_e$ vs.
stellar mass $M_\star$. Indeed, assuming that about half of the initial
baryonic mass is in form of stars, the system would lie initially a factor
$\simeq 2.5$ below the local mass-size relationship for ETGs. The initial
(i.e.\ before mass loss and expansion, Eq.\ \ref{eq:tdyn}) dynamical time
is $t_{\rm dyn}\approx 5$ Myr. Note that a smaller initial size would
shorten the dynamical time, exacerbating the problems pointed out in
Section \ref{sec:discussion}.

In summary, the parameters affecting the results of our simulations are
the ratio of mass between the total and baryonic components $M_{\rm
vir}/M_{\rm B(t=0)}$; the corresponding ratio of scale-lengths $R_{\rm
vir}/a$; the fraction of baryonic mass lost $(1-\epsilon)$, and the time
$\Delta t$ over which the loss occurs. We performed simulations covering
broad ranges of the latter two quantities, while in most runs we kept the
former two at the fiducial values discussed above. We checked however that
none of our qualitative conclusion is affected by factor $\sim 2$
variations of them, and likely even by larger ones (see discussion at the
end of Section \ref{sec:results}).

\begin{figure}
  \centerline{\includegraphics[width=9cm]{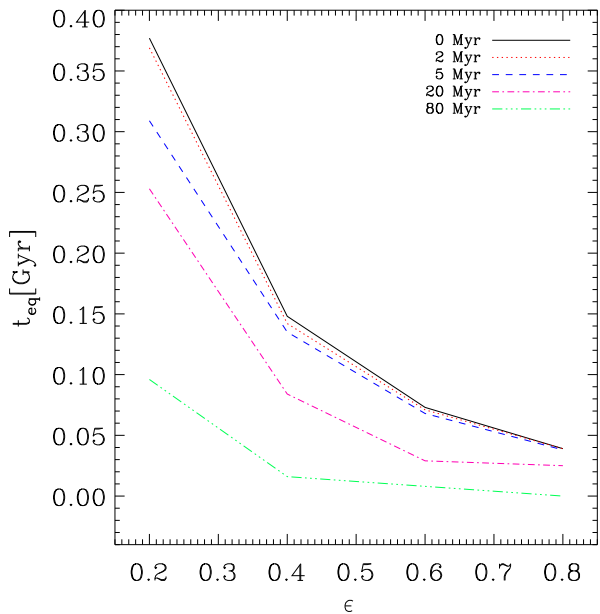}}
  \caption{The time needed, after the end of mass loss, to recover an equilibrium configuration as
  a function of the parameter $\epsilon\equiv
M_{B,fin}/M_{B,ini}$, for different ejection times $\Delta t$.
  This time is defined by the epoch after which the size never changes more than 10\% with respect
  to the final value.
} \label{fig:teq}
\end{figure}

\section{Results}
\label{sec:results}

As a preliminary sanity check, we ran simulations without DM with ejection
times $\Delta t=0$ Myr and $80$ Myr, i.e., much shorter or much longer
than $t_{\rm dyn}$, respectively. The ratio of the  final to the initial
half-mass size $R_{\rm f}/R_{\rm i}$ generally agrees well with the
expectations given by Eq.~(1) and (2) (see Fig.~\ref{fig:sizinc}).
However, when the mass loss is fast and approaches 50\%, the analytical
formula Eq.~(1) increasingly over-predicts the numerical result,
underlying the well known fact that the divergence does not occur at
$\epsilon=0.5$ in numerical simulations. This is in keeping with previous
findings (e.g.\ Geyer \& Burkert 2001).

We now turn to cases with DM included. Fig.~\ref{fig:sizinc} illustrates
the size expansion as a function of the fraction of remaining baryonic
mass $\epsilon= 0.2$, $0.4$, $0.6$, and $0.8$, for different ejection
times $\Delta t=0$, $2$, $5$, $20$, and $80$ Myr. The expansion increases
with decreasing $\epsilon$ and $\Delta t$, but it is milder with respect
to the corresponding purely baryonic case. In particular, the system is no
longer disrupted even when the ejection is impulsive ($\Delta t=0$) and
$\epsilon$ is as low as =0.2; this is expected since DM constitutes the
dominant source of gravitational potential at large radii. We have
verified, with a few sample runs, that the case with $\Delta t=80$ Myr is
representative of the "slow expulsion" regime $\Delta t \gg t_{\rm dyn}$.
In other words, the expansion factor do not decrease any more for larger
values of $\Delta t$. Note also that, as already mentioned, $\Delta t=20$
is likely the case that best approximates, among those shown, the QSO
driven gas expulsion predicted by the G04 model (see
Fig.~\ref{fig:modelo}), and considered by Fan et al.\ (2008; 2010). The
corresponding expansion is significantly smaller than that achieved for
instantaneous ejection.

Fig.~\ref{fig:rad_t} shows in detail the time evolution of the system size
during and after the ejection. For sake of comparison we include also a
case without DM. For $\Delta t\gg t_{\rm dyn}$, the size increases mostly
during the mass loss, and stabilizes soon after $\Delta t$, so that the
system is actually in quasi-equilibrium during the ejection process;
contrariwise, for impulsive ejection with $\Delta t \ll t_{\rm dyn}$ the
initial equilibrium is totally broken, and the system expands
significantly more. The size undergoes damped oscillations before
stabilizing, more important for smaller $\epsilon$ or $\Delta t$.
Fig.~\ref{fig:teq} shows the time needed, after the end of mass loss, to
recover a substantially stable configuration. This turns out to be shorter
than the corresponding time without the DM component, as expected on an
intuitive basis, due to the stabilizing effect of the latter (for
instance, compare the thin solid curve in the top-right panel of Fig.
\ref{fig:rad_t}, with the thick solid curve in the bottom-left panel).

\begin{figure*}
\centerline{\includegraphics[width=9cm]{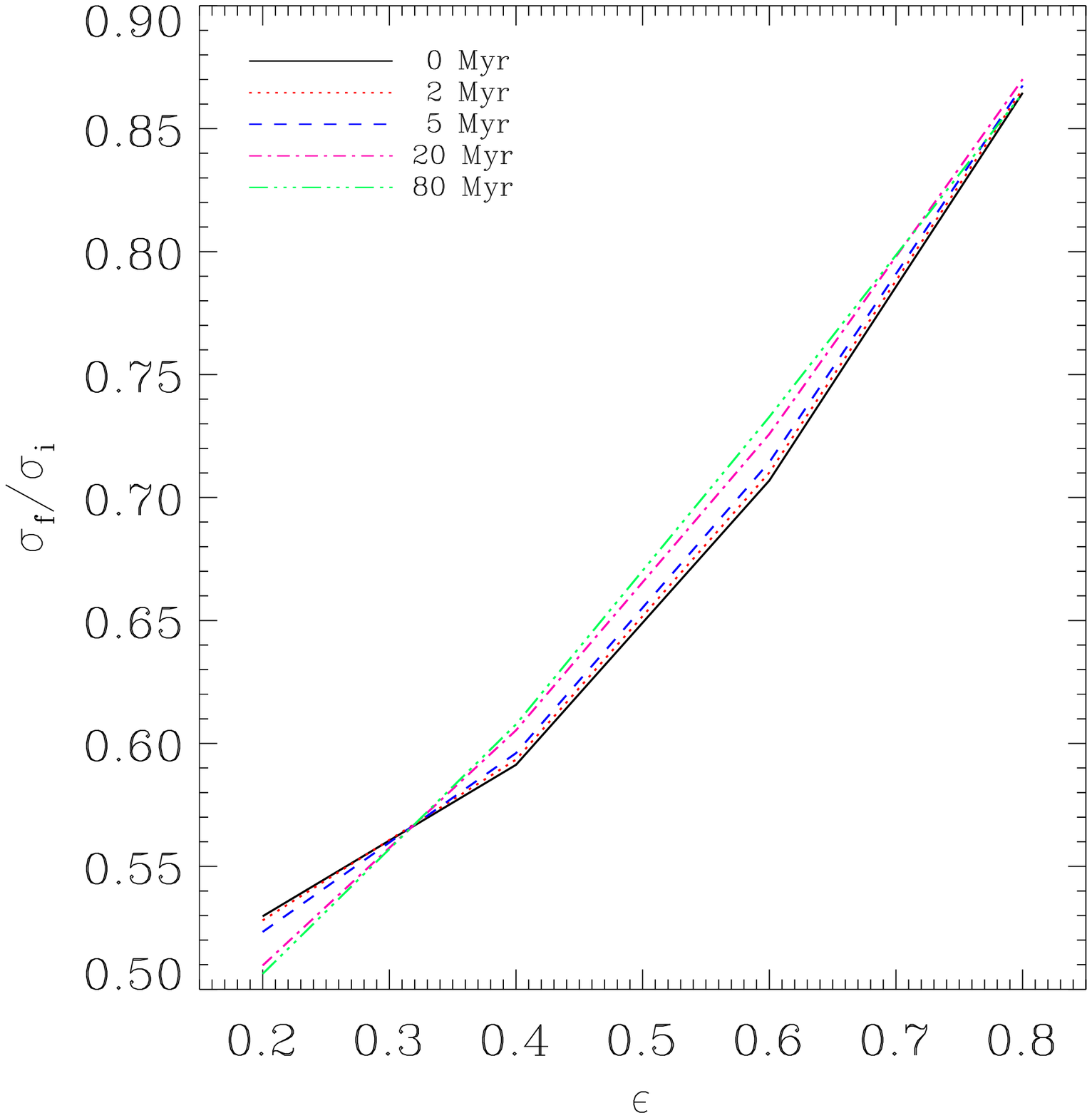}\includegraphics[width=9cm]{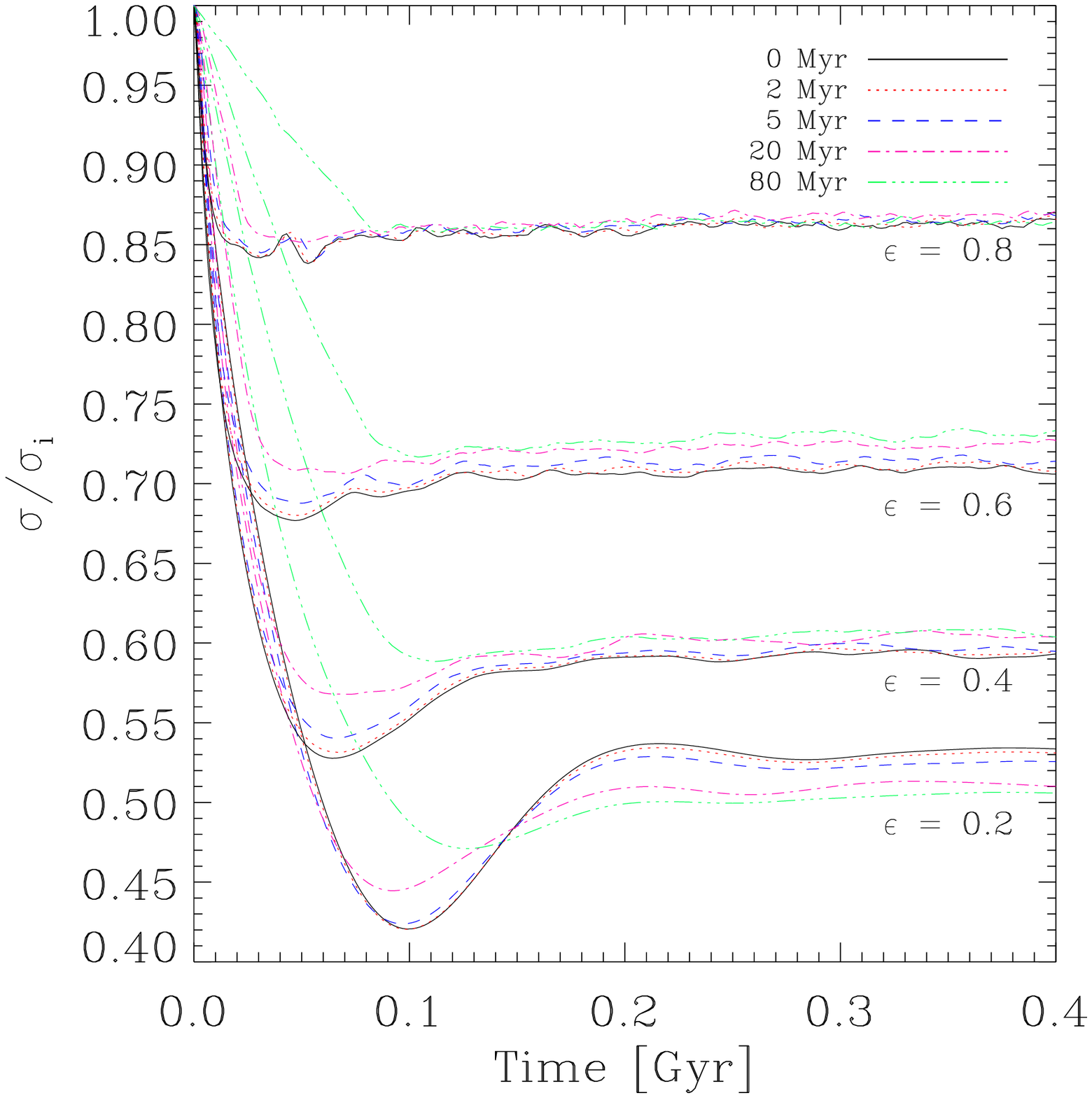}}
\caption{Left panel: Ratio $\sigma_{\rm f}/\sigma_{\rm i}$ of the final to
initial mean 3D stellar velocity dispersion as a function of $\epsilon$,
for different ejection times $\Delta t$. Right panel: same as a function
of time, for different values of the diet parameter $\epsilon$ (as
labeled) and of the ejection times $\Delta t$.} \label{fig:sig}
\end{figure*}

Fig.~\ref{fig:sig} shows the evolution of the average stellar velocity
dispersion. During the size expansion the system cools down and the
stellar random motions slow, reducing the average dispersion. The net
effect is found to be much evident for strong ejection (small $\epsilon$)
but almost independent of the ejection time $\Delta t$.

\begin{figure*}
  \centerline{\includegraphics[width=9cm]{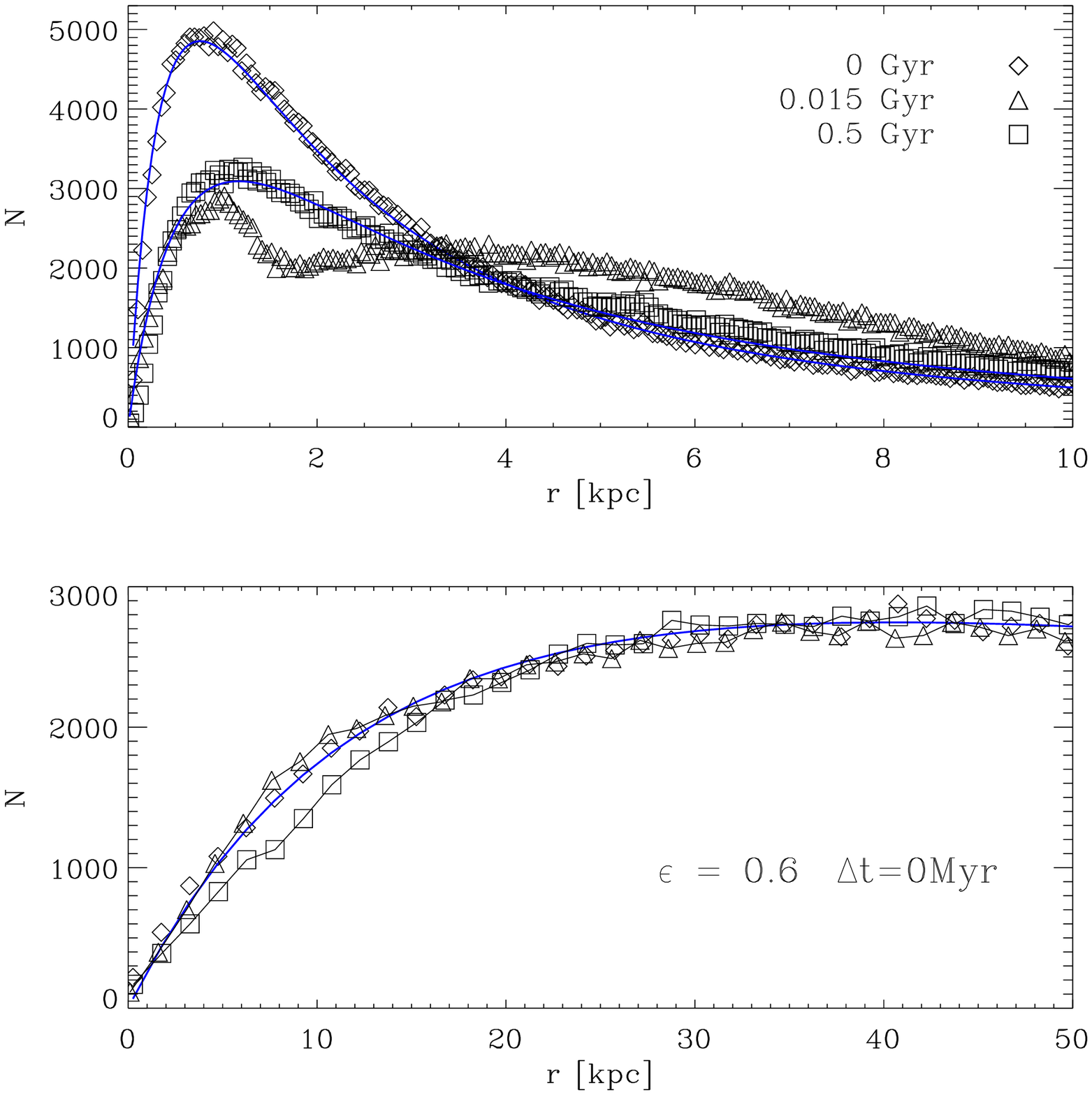}
  \includegraphics[width=9cm]{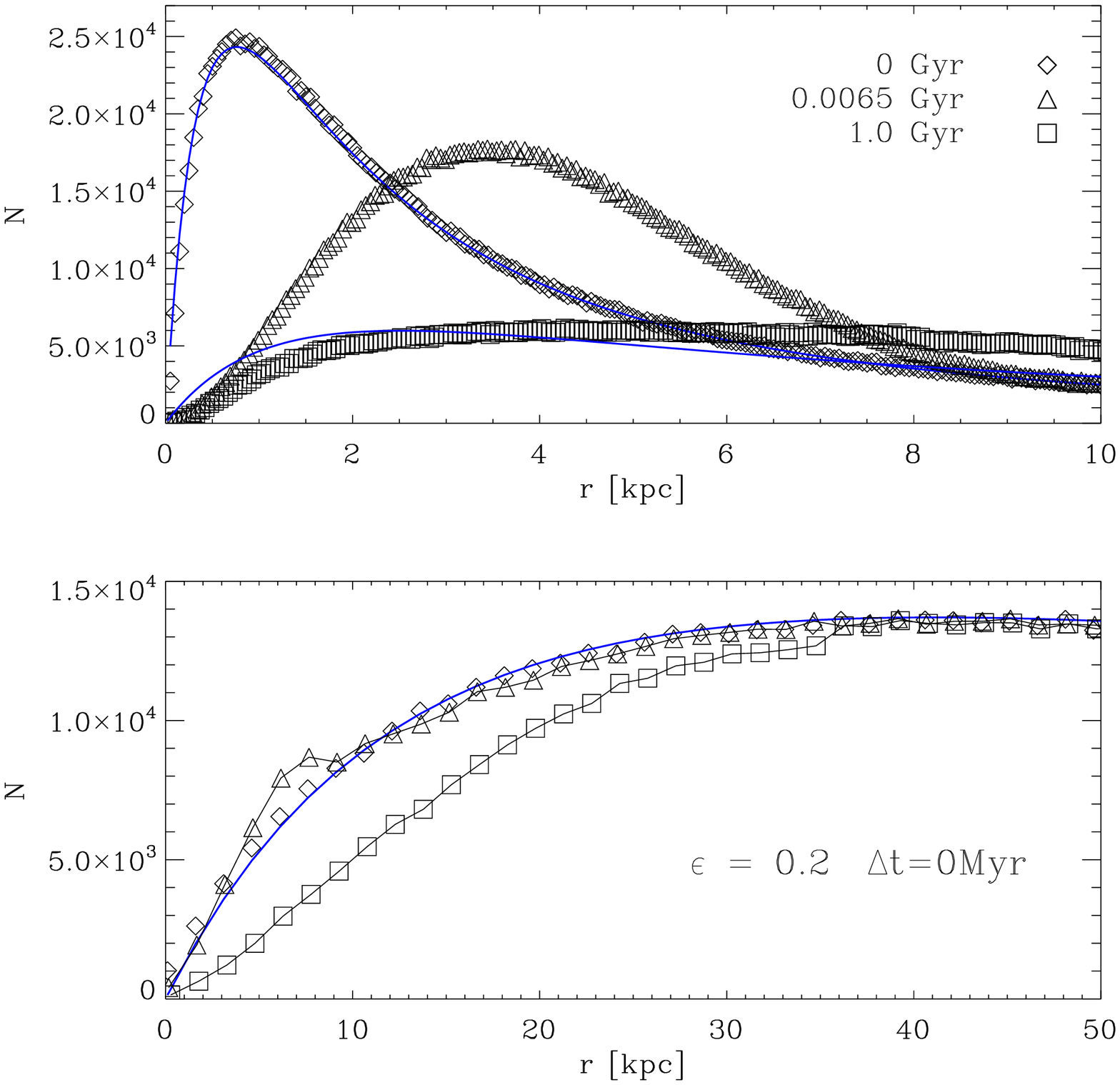}}
  \caption{Two examples of the effects of baryon expulsion on
the density profiles (shown as radial distribution of particles) of
baryons (top panels) and DM  (bottom panels, showing only the inner
region). The left column refers to $\epsilon=0.6$ and $\Delta t=0$, while
the right one is for $\epsilon=0.2$ and $\Delta t=0$. We plot three
snapshots, i.e.\ the initial one, one wherein the baryons are far from
equilibrium, and the final one (times indicated in the bottom panels). The
blue solid lines are $\chi^2$ fits to the distributions with Hernquist and
NFW profiles respectively, which we show in the former case only for the
initial and final states, and in the latter only for the inital one. }
\label{fig:perfiles}
\end{figure*}

In Fig~\ref{fig:perfiles} we illustrate the effects of baryon expulsion on
the density profiles of both components. For $\epsilon \ge 0.6$ and fast
expulsion, the baryonic component, after a violently disturbed phase,
eventually recovers a density distribution  reasonably well described by
the Hernquist formula, albeit with a larger scalelength (for instance
$a=2.34$ with reduced $\chi^2=2.6$ for $\epsilon = 0.6$, shown in the
figure). By converse, at lower $\epsilon$, the Hernquist fit becomes
increasingly unacceptable for the final equilibrium profile (e.g.\ $a=5$
and reduced $\chi^2=41$ for $\epsilon = 0.2$). This is not surprising,
since the final bound state is increasingly dictated by the presence of
the embedding DM halo. Actually for $\epsilon \lesssim 0.5$ the system
would dissolve if not for the halo. In any case slower expulsion with same
$\epsilon$ yields lower deviations from the Hernquist functional form. In
particular, for $\epsilon \ge 0.6$ and $\Delta t=80$ Myr the Hernquist
formula provides a good description of the density distribution even
during the mass loss-expansion phase.


For DM, on large scales the profile is unaffected, while in the inner
region the baryon expansion drags an expansion of the DM particles. As a
result, the DM profile in the galactic region is always flattened to some
level with respect to the original NFW shape.

\begin{figure}
  \centerline{\includegraphics[width=9cm,height=5cm]{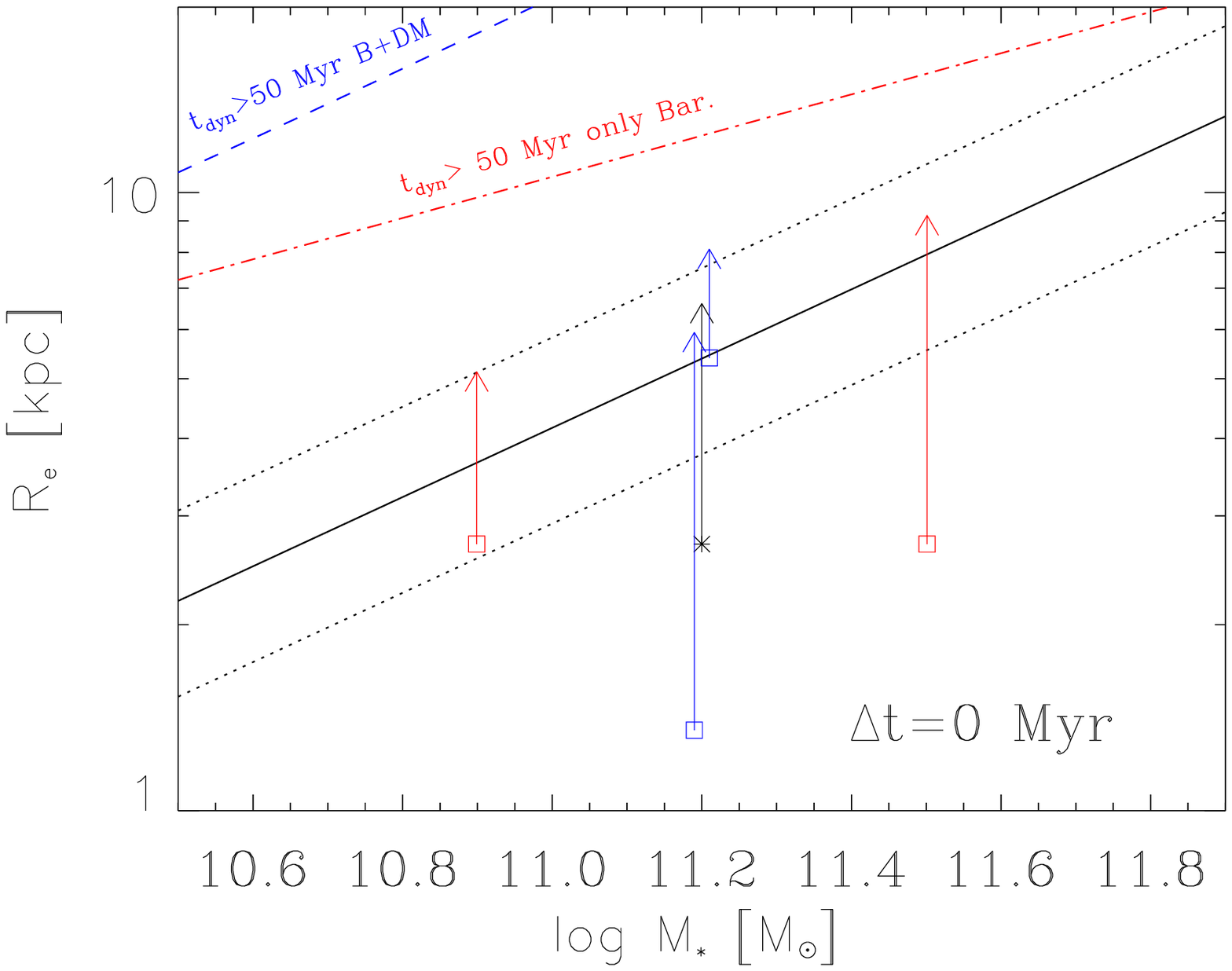}}
  \centerline{\includegraphics[width=9cm,height=5cm]{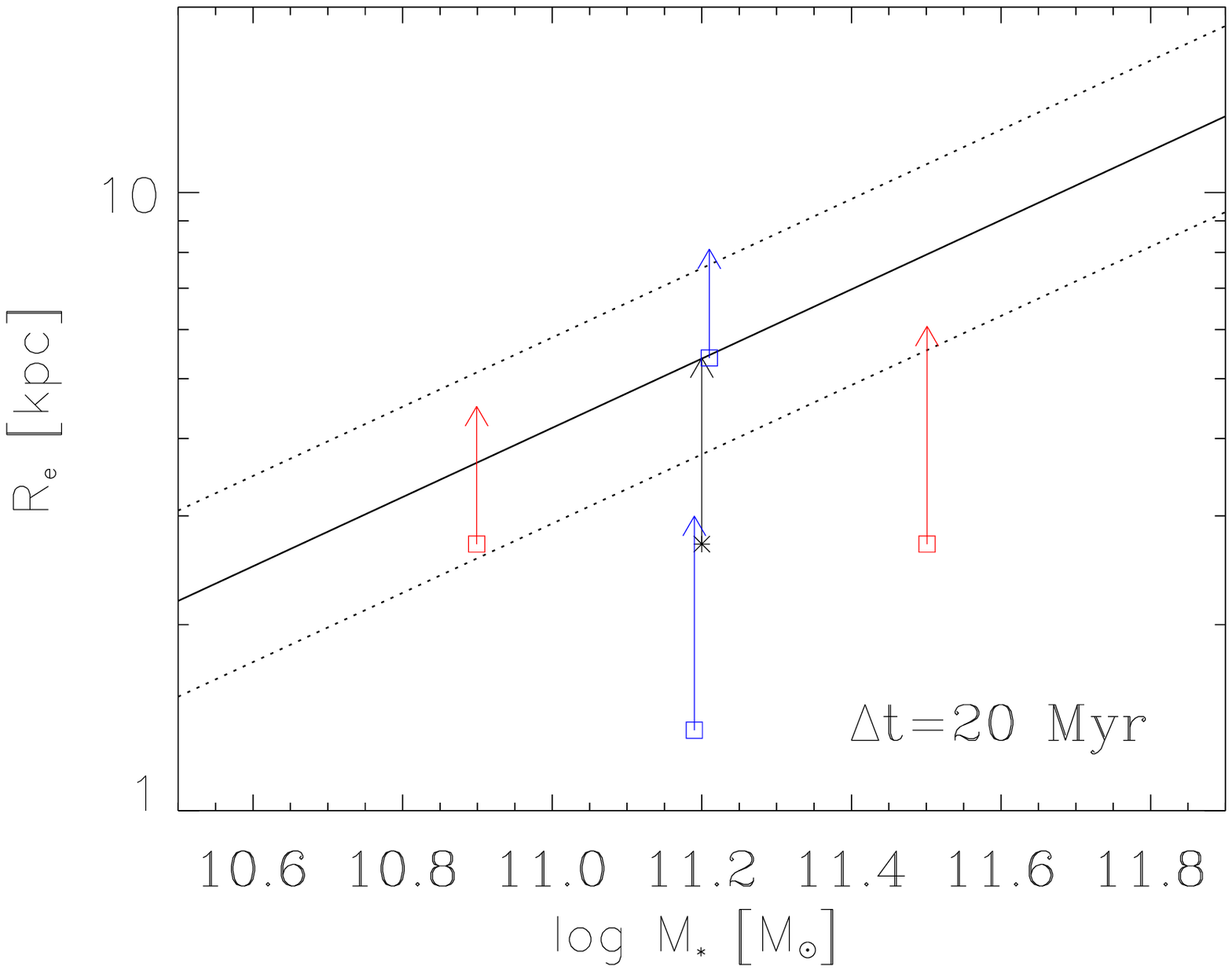}}
  \centerline{\includegraphics[width=9cm,height=5cm]{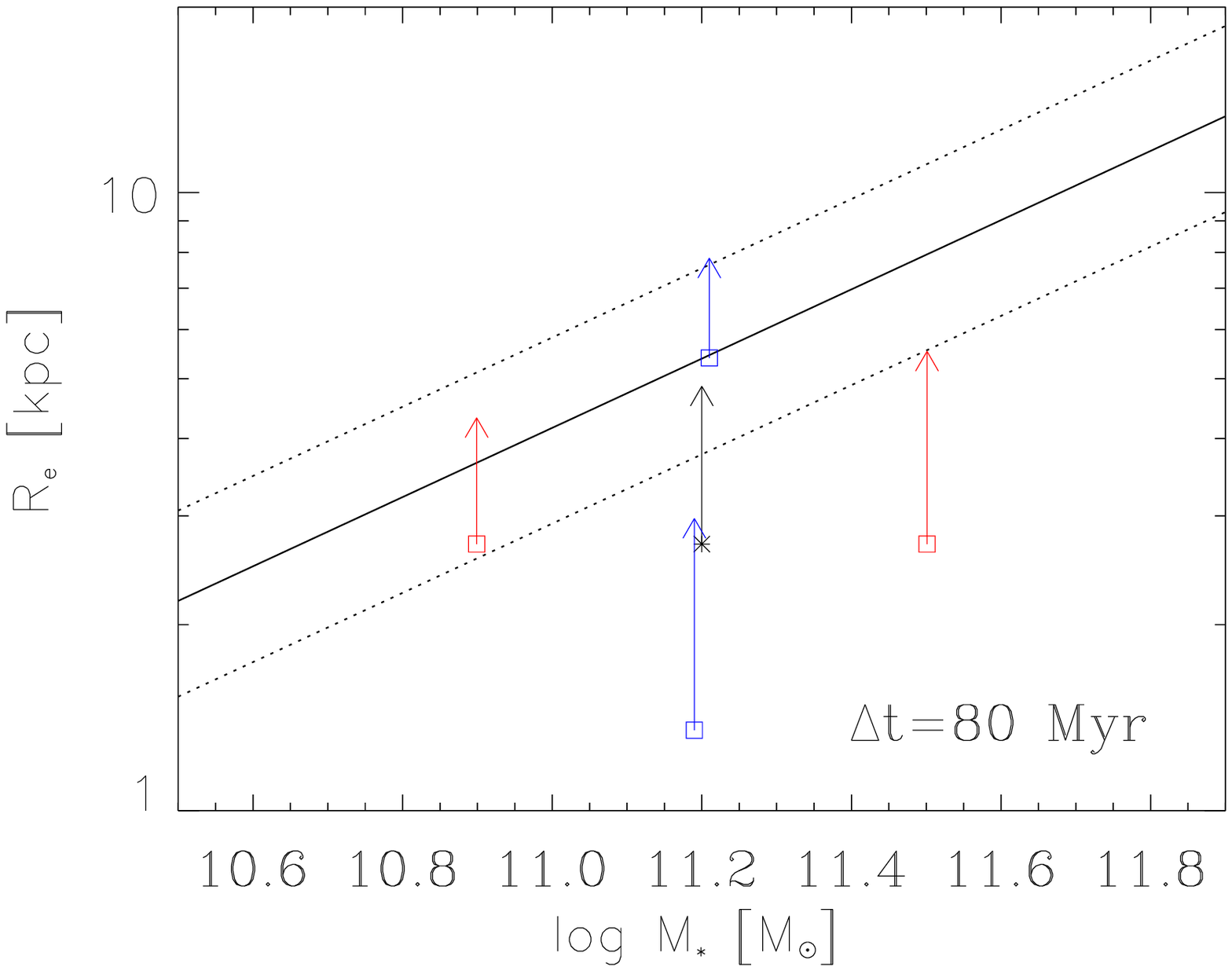}}
  \caption{Initial position (points) of our runs in the stellar mass-size plane,
  and its evolution (arrows) due to gas ejection amounting to 60\% of the initial baryonic mass (i.e.\ for
  $\epsilon\equiv M_{B,fin}/M_{B,ini}=0.4$). The various panels refer to different expulsion times $\Delta
  t$,
  as indicated. The starred point with black arrow represents the reference model,
  while the other four points-arrows show the behaviour of systems having an initial baryonic mass or
  radius
  twice smaller or greater (while the DM halo remains identical).
  For ease of reading, we have artificially displaced slightly in mass the points
  corresponding to variations in radius. The solid diagonal line is the mass-size relationship
  for local ellipticals (Shen et al. 2003), with the 1 $\sigma$ dispersion depicted by dotted lines.
  The upper panel shows the lines above which the initial dynamical
  time is $\gtrsim 50$ Myr, without DM (dot-dashed, Eq. \ref{eq:tdynapp}) or including it (dashed, Eq.
  \ref{eq:tdyn})
  (see Section \ref{sec:discussion}).
  } \label{fig:var_par}
\end{figure}

Fig~\ref{fig:var_par} shows the sensitivity of our results to the
parameters of the initial baryonic configuration. As expected, the effects
are in the sense that, whenever the DM contributes more (less) to the mass
inside the region occupied by the baryonic  system, the latter expands
less (more). This may occur by increasing (decreasing) the ratios $M_{\rm
vir}/M_{\rm B(t=0)}$ or $a/R_{\rm vir}$. Note the trade-off between
variations in initial size and its amplification due to mass loss. As a
result, the final size is relatively insensitive to the initial one,
particularly for fast expulsion. For instance, when the impulsive mass
loss is 60\% ($\epsilon=0.4$), a factor 4 change of the adopted initial
$R_{e}$, yields only a factor $\sim 1.5$ change in the final $R_{e}$ (blue
points and arrows in Fig~\ref{fig:var_par}). This could have a role in
explaining the relatively low scatter of the observed local mass-size
relationship. {\rev Also, an increase (decrease) of $M_{\rm B(t=0)}$,
keeping fixed all other parameters, and in particular $M_{\rm vir}$,
yields a larger (smaller) expansion. This tends to keep the expansion
close to that required to maintain  the final state of the system in the
region of local mass-size relation (red points and arrows in
Fig~\ref{fig:var_par}).}

\begin{figure}
  \centerline{\includegraphics[width=9cm]{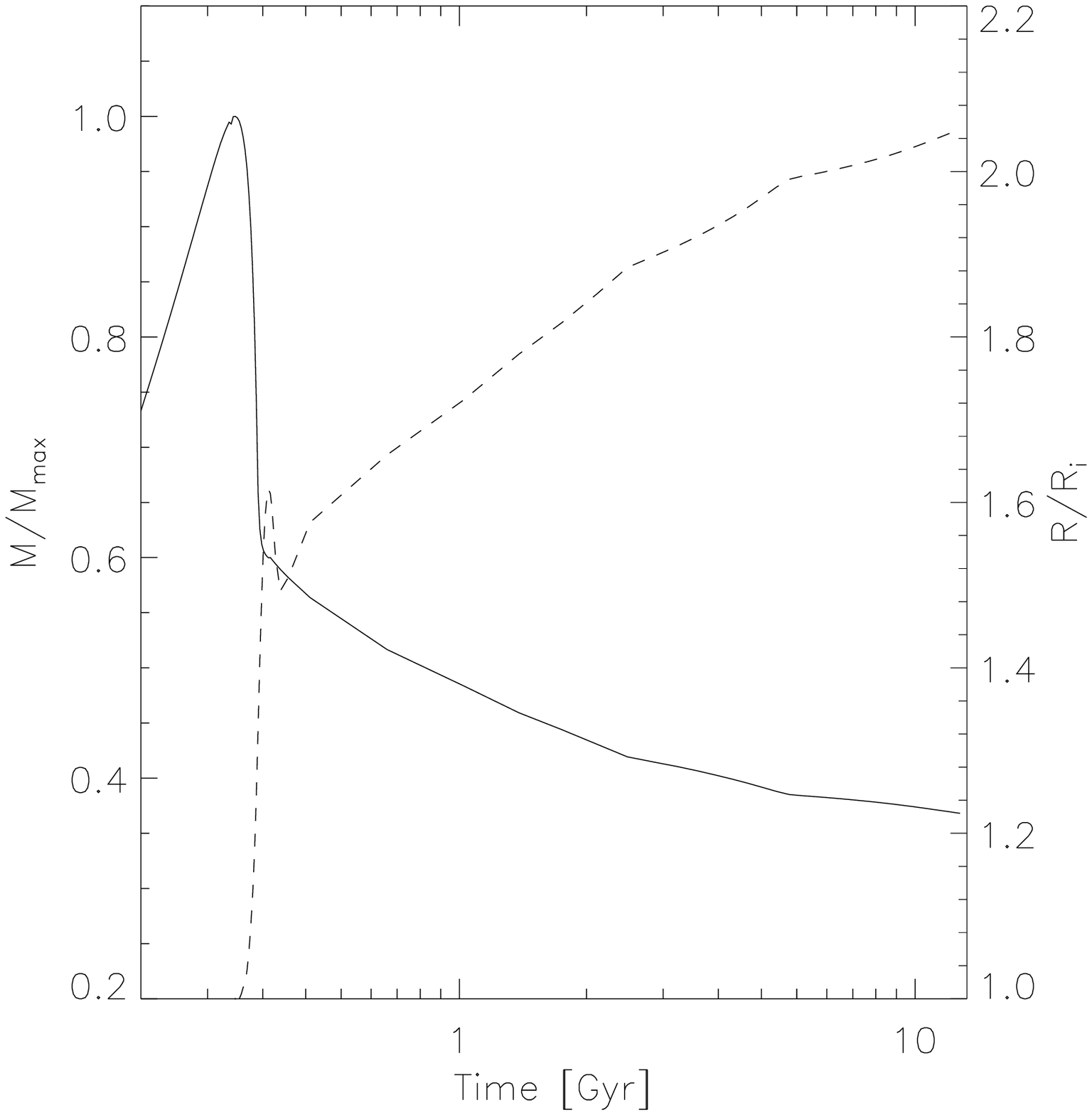}}
  \caption{Example of evolution of the total baryonic mass (star forming gas+stars) in the
  G04 model for co-evolution of SMBH and spheroids (solid line, left axis), and the
  increase in size (dashed line, right axis) according to our corresponding simulation. {\rev The mass
  is measured in units of $M_{max}$, which is the maximum value of the total baryonic mass
  (stars+cold gas) attained during its formation. This is reached just before the galactic winds
  eject the cold gas and stop further star formation.} The abrupt decrease of the mass after $\sim 0.3$
  Gyr
  marks the ejection of the cold gas, not yet converted into stars, by the AGN-driven wind, while the
  subsequent
  slow decrease is due to stellar mass returned to the ISM, under the assumption that
  the galaxy potential cannot retain it. See text for more details.
  } \label{fig:modelo}
\end{figure}

\section{Discussion and conclusions}
\label{sec:discussion}

An inspection to the previous figures, in particular to Fig.\
\ref{fig:rad_t}  (and to a lesser extent Fig.\ \ref{fig:teq}), reveals the
main problem to explain the {\it observed} size evolution of ETGs with the
puffing-up scenario. On one hand, our simulations confirm that, even in
presence of a DM component, a factor $\sim 2$ increase in size can be
expected in any galaxy formation model in which the spheroid loses $\sim
50$\% of its baryonic mass, in particular when this happens on a timescale
of the order of the dynamical timescale. Moreover, the expansion is larger
for systems initially more compacts, resulting in an interesting self
regulation of the final size (Fig. \ref{fig:var_par}). However, if this
mass is constituted by the star forming gas, in scenarios in which a
galactic wind suddenly sterilizes the galaxy (such as in the G04 model
considered by Fan et al.\ 2008, 2010), the puffing-up occurs far too close
to the last episode of star formation. Indeed, the galaxy is predicted to
be smaller than the final size only for a very short time after expulsion,
less than a few dynamical times, i.e.\ less than $\sim 20- 30$ Myr for the
adopted initial configuration. This is at least a factor 20 less than the
estimated ages of stellar populations in high-z compact galaxies ($\gtrsim
0.5-1$ Gyr; e.g.\ Longhetti et al 2007; Damjanov et al.\ 2009). Even
taking into account generous uncertainties of these estimated ages, it
seems clear that a substantial contribution of galactic winds to the size
evolution observed so far can be safely ruled out. Nevertheless, this
process should have had a role in deciding the size of ETGs, if
(QSO-driven) galactic winds caused their sudden death, but its signature
should be searched for in much younger systems. This poses huge
observational challenges with present facilities.

Since the expansion timescale is proportional to the dynamical time
$t_{\rm dyn}\propto M^{-0.5} R^{1.5}$, it could be suspected that the
problem originates from our choice of initial conditions, such as an
excessive initial baryonic mass or an insufficient initial radius. However
to get a factor $\gtrsim 10$ increase (a minimal requirement) of the
dynamical time, the initial size should be increased at least by a factor
$\gtrsim 5$, or the mass decreased by a factor $\gtrsim 100$. Actually,
even these large changes of the baryonic distribution would not suffice,
since in both cases the fixed DM component would become important to set
the dynamical time in the galactic region (Eq.\ \ref{eq:tdyn}). Anyway, in
both cases the initial state of the system would already lie well above
the observed {\it local} size-mass relationship. This is illustrated by
lines in the top panel of Fig.~\ref{fig:var_par}, delimitating the region
of the plane $R_{e}-M_{*}$ where $t_{\rm dyn} \gtrsim 50$ Myr. As can be
appreciated there, the inclusion of a DM contribution to estimate $t_{\rm
dyn}$ makes the argument even stronger. {\rev Moreover, our standard
initial conditions put the system about a factor 2.5 below the z=0
size-mass relationship, if about half of the baryons are in stars (in
other word if $\epsilon \sim 0.5$). A non negligible fraction of compact
high-z ETGs are found up to a factor $\sim 3-5$ below this relationship
(e.g.\ Zirm et al. 2007, Toft et al. 2007, Van Dokkum et al. 2008, Cimatti
et al.\ 2008, McGrath et al.\ 2008, Cassata et al.\ 2010). For these, the
initial dynamical time is even shorter, exacerbating the problem, as we
anticipated in Section \ref{sec:method}. On the other hand, given the
anti-correlation between initial size and expansion $R_f/R_i$ achieved
after mass loss, discussed at the end of Section \ref{sec:results} and
shown in Fig.\ \ref{fig:var_par}, also for these extremely compact objects
it is conceivable to get final sizes close to the locally observed ones.}

We have also verified, with a few sample simulations, that this general
conclusion on the shortness of the expansion timescale is not
significantly affected by assumptions such as that the gas and stars share
the same profile before ejection of the former, or that this ejection
occurs homogenously in the system. The same holds true for different
choices of the density profile {\rev of the DM and baryonic} components
(Eq. \ref{eq:nfw} and Eq. \ref{eq:her}), provided they describe in a
reasonable manner the density distributions of the respective mass
distribution.

Even in the case of expansion driven by stellar mass loss {\rev the
problem of the excessive shortness of the expansion timescale} is likely
important, though less clear cut. Indeed, Damjanov et al.\ 2009 pointed
out it, basing their reasoning on the star cluster approximation reviewed
in Section \ref{sec:cluster}, and in the performing simplified estimates
of mass loss due to stellar evolution. They found that the fraction of
mass lost during the passive evolution of stellar populations can be as
large as 30-50\%, depending on the IMF, but {\rev the majority} of this
loss occurs in less than 0.5 Gyr (see their figure 7). This timescale is
still younger than the typical estimated ages of high redshift compact
ETGs. However, it should be pointed out that the details of this result
depend also on the adopted recipes for stellar lifetimes and yields. These
ingredients have some uncertainty (e.g.\ Romano et al.\ 2005). As a
result, it cannot be totally excluded that passively evolving ETGs lost
20-30\% of the residual baryonic mass even $\sim 0.5$ Gyr after the end of
their main star forming phase. This would produce a small (due also to the
relatively inefficient size increasing effect of slow mass loss, Figs.\
\ref{fig:sizinc} and \ref{fig:rad_t}), but not negligible, contribution to
the claimed size evolution. Moreover, in this case the uncertainty on the
relatively difficult estimation of ages could have some importance, at
least for the youngest observed high-z compact ETGs.

To better illustrate the above points, we show in Fig. \ref{fig:modelo}
the result of a sample simulation, applied to a specific semi-analytic
galaxy formation model including both processes, namely QSO driven
galactic wind and mass loss from stars due to stellar evolution. {\rev The
figure displays} the time evolution of baryonic mass as predicted by the
G04 spheroid-SMBH co-evolution model (with the parameters as in Lapi et
al.\ 2006), in a $10^{13} M_\odot$ DM halo that virializes at $z=4$,
together with the corresponding increase in size, computed with the
procedure described in the present paper. The abrupt decrease of the mass
after $\sim 0.3$ Gyr marks the ejection of gas by the AGN-driven wind.
Note that, though this is a relatively fast process,  it still does occur,
according to the adopted recipes, on a timescales of a few $t_{\rm dyn}$.
This holds true over a wide range of model parameters. As a consequence,
the corresponding puffing-up is milder (a factor $\sim 1.5$) than the
maximal one (for a given $\epsilon$), achieved when $\Delta t=0$. The
subsequent slow decrease of mass and moderate increase in size (a further
factor  $\sim 1.35$, for a total expansion of $\sim 2$), is due to stellar
mass returned to the ISM, under the assumption that the galaxy potential
cannot retain it. The size expansion achieved after the epoch in which
stellar populations are older than $\sim 0.5-1$ Gyr is 25\% - 20\%,

In conclusion, the putative puffing up related to large scale galactic
winds, quickly ejecting a substantial fraction of baryonic mass, can be an
important phenomenon, but is still not observed. {\rev In particular, it
cannot be invoked to explain the size evolution of ETG from $z\simeq 2.5$
to $z=0$ observed in the presently available data}. By converse, the
secular adiabatic expansion, related to the mass returned to the ISM by
stars during the final stages of their evolution, could contribute, but
not dominate, the observed size evolution of ETGs. Nevertheless, it is
relevant to further investigate also this contribution, since it seems
that none of the processes or biases considered so far can explain alone
this evolution (e.g.\ Hopkins et al.\ 2010).

\section*{Acknowledgments}
C.R-F.\ and G.L.G.\ acknowledge warm hospitality by INAF-Trieste and
IATE-C\'ordoba, respectively, during the development of the present work.

This work has been partially supported by the Consejo de Investigaciones
Cient\'{\i}ficas y T\'ecnicas de la Rep\'ublica Argentina (CONICET) and
the Secretar\'{\i}a de Ciencia y T\'ecnica de la Universidad Nacional de
C\'ordoba (SeCyT).

We thank Giuseppe Murante for carefully reading the manuscript and for
useful suggestions, {\rev and the anonymous referee for proposing several
improvements of the paper.}

{}

\clearpage

\end{document}